\documentclass[a4paper,12pt]{article}

\usepackage{latexsym}
\usepackage{amsmath}	%%% AMS Package
\usepackage{amssymb}	%%% AMS Package
\usepackage{bm}
\usepackage{graphicx}
\usepackage{wrapfig}
\usepackage{fancybox}
\usepackage[subrefformat=parens]{subcaption}
\newcommand{\bea}{\begin{eqnarray}}
\newcommand{\eea}{\end{eqnarray}}
\begin{document}
\begin{titlepage}
%\begin{flushright}
%AAA-BB-111 \\
%\end{flushright}
%
%\vspace*{10mm}
\begin{center}
\baselineskip 25pt 
{\Large\bf
%%%%%%%%%%%%%%%%%%%%%%%%%%%%%%%%%%%%%%%%%%%%%%%%%%%%%%%%%%%%%%%
R-parity Conserving Minimal SUSY U(1)$_{X}$ Model
%%%%%%%%%%%%%%%%%%%%%%%%%%%%%%%%%%%%%%%%%%%%%%%%%%%%%%%%%%%%%%%
}
\end{center}
\vspace{2mm}
\begin{center}
{\large
Satsuki Oda$^{~a,b}$ \footnote{satsuki.oda@oist.jp},
Nobuchika Okada$^{~c}$ \footnote{okadan@ua.edu}, 
Nathan Papapietro$^{~c}$, \\
and 
Dai-suke Takahashi$^{~a,b}$ \footnote{darksusy@gmail.com}
}
\end{center}
\vspace{2mm}

\begin{center}
{\it
$^{a}$Okinawa Institute of Science and Technology Graduate University (OIST), \\ 
Onna, Okinawa 904-0495, Japan
\vspace{5mm}

$^{b}$Institute for Pacific Rim Studies, Meio University, \\
Nago, Okinawa 905-8585, Japan \\
\vspace{5mm}

$^{c}$Department of Physics and Astronomy, University of Alabama, \\
Tuscaloosa, Alabama 35487, USA
}
\end{center}
\vspace{3mm}
%%%%%%%%%%
\begin{abstract}
%%%%%%%%%%

We propose a minimal gauged U(1)$_X$ extension of the Minimal Supersymmetric (SUSY) Standard Model (MSSM) 
  with R-parity conservation. 
In this model, U(1)$_X$ is a generalization of the well-known U(1) $B-L$ (baryon number minus lepton number).
Apart from the MSSM particle content, the model includes three right-handed neutrino (RHN) chiral superfields, 
  each carrying a unit U(1)$_X$ charge. 
In the presence of RHNs, the model is free from all gauge and mixed gauge-gravitational anomalies. 
However, there are no U(1)$_X$ Higgs chiral superfields with U(1)$_X$ charge $\pm2$ involved in the model.
Two of the RHN superfields are assigned an odd R-parity, while the last one ($\Psi$) has an even parity. 
The U(1)$_X$ symmetry is radiatively broken by the vacuum expectation value (VEV) 
  of the scalar component of $\Psi$ after renormalization group evolution of the scalar potential.
As a consequence of the absence of U(1)$_X$ Higgs fields and the novel R-parity assignment, 
  the three light neutrinos consist of one massless neutrino and two Dirac neutrinos.
In the early universe, the right-handed components of the Dirac neutrinos are in thermal equilibrium
  with the Standard Model (SM) particles through the U(1)$_X$ gauge ($Z^\prime$) boson. 
The extra energy density from the right-handed neutrinos is constrained 
  to avoid disrupting the success of Big Bang Nucleosynthesis (BBN), 
  leading to a lower bound on the scale of U(1)$_X$ symmetry breaking.
In our model, a mixture of the U(1)$_X$ gaugino and the fermionic component of $\Psi$ 
   becomes a new dark matter (DM) candidate if it is the lightest sparticle mass eigenstate. 
We examine this DM phenomenology and identify a parameter region that reproduces the observed DM relic density.
Furthermore, we consider constraints from the search for $Z'$ boson resonance at the Large Hadron Collider (LHC). 
The three constraints obtained from the success of BBN, the observed DM relic density, and the $Z^\prime$ resonance 
   search at the LHC complement each other, narrowing down the allowed parameter region. 
Additionally, we note that once the $Z^\prime$ boson is discovered at the LHC, 
  it can serve as a probe to determine the number of left-handed plus right-handed neutrinos
  through measurements of its invisible decay width.

\end{abstract}
\end{titlepage}

%%%%%%%%%%%%%%%%%%%%%%%%%%%%%%
\newpage
\section{Introduction}
%%%%%%%%%%%%%%%%%%%%%%%%%%%%%%

The $B-L$ (baryon number minus lepton number)  
  is the unique anomaly-free global U(1) symmetry in the Standard Model (SM). 
We can easily gauge this global symmetry, and the so-called minimal $B-L$ model~\cite{MBL1}-\cite{MBL6} 
  is a simple gauged $B-L$ extension of the Standard Model (SM), where three right-handed neutrinos (RHNs) 
  and an SM gauge-singlet Higgs field carrying two units of the $B-L$ charge are introduced. 
The model is free from all gauge and mixed gauge-gravitational anomalies in the presence of the RHNs.  
Associated with a $B-L$ symmetry breaking by a vacuum expectation value (VEV) 
  of the $B-L$ Higgs field, the $B-L$ gauge ($Z^\prime$) boson and the RHNs acquire their masses. 
After the electroweak symmetry breaking, the seesaw mechanism~\cite{seesaw1}-\cite{seesaw5} works 
  to generate tiny SM neutrino masses naturally.

Although the scale of the U(1)$_{B-L}$ gauge symmetry breaking is arbitrary as long as it is phenomenologically available, 
  the breaking scale of $1-10$ TeV is probably of the best interest in the viewpoint of the Large Hadron Collider (LHC) experiment. 
However, the argument of the gauge hierarchy problem of the SM is also applicable to the $B-L$ Higgs field
  and quantum corrections to the $B-L$ Higgs field mass squared are quadratically sensitive to the scale of ultraviolet (UV) theory that we expect to take place at high energies. 
Thus, the $B-L$ symmetry breaking scale is unstable against quantum corrections.   
As is well-known, supersymmetric (SUSY) extension is one of the most promising ways to make the theory UV insensitive 
  and ensure the stability of the symmetry breaking scale. 
Interestingly, we can see that in the SUSY extension of the minimal $B-L$ extended SM, 
  the $B-L$ asymmetry is also radiatively broken in the presence of the soft SUSY breaking mass terms 
  at a high energy \cite{Khalil:2007dr, FileviezPerez:2010ek, Burell:2011wh}.  
In this radiative symmetry breaking, the soft SUSY breaking parameters
  control the $B-L$ symmetry breaking scale, and hence the scale of $B-L$ symmetry breaking
  lies at the TeV scale from the naturalness point of view.

SUSY extension of the $B-L$ model opens a new interesting possibility. 
As has been proposed in Ref.~\cite{Barger:2008wn}, 
  it is unnecessary to introduce the $B-L$ Higgs field in this extension 
  since the scalar partner of a RHN chiral superfield can play the role of breaking the $B-L$ gauge symmetry.  
Therefore, we may define the ``minimal SUSY $B-L$ model'' by excluding the $B-L$ Higgs chiral superfields
  carrying $B-L$ charge $\pm 2$.
Interestingly, such a particle content can be derived from heterotic strings with flux compactifications \cite{Braun:2005nv, Anderson:2009mh}. 
In Ref.~\cite{Barger:2008wn}, a negative soft mass squared for a right-handed sneutrino 
  is assumed to break the $B-L$ gauge symmetry, so that the $B-L$ symmetry 
  breaking occurs at the TeV scale.  
Associated with this symmetry breaking, R-parity is also spontaneously broken. 
See, for example, Refs.~\cite{FileviezPerez:2009gr,  Everett:2009vy, FileviezPerez:2012mj, Marshall:2014kea} 
  for interesting phenomenology with the R-parity violation. 
Through the non-zero VEV of the right-handed sneutrino, neutrinos have mass mixings  
  with MSSM Higgsinos, MSSM neutralinos, and $B-L$ gaugino. 
Although the neutrino mass matrix becomes very complicated, 
  it has enough degrees of freedom to reproduce the neutrino oscillation data 
  with a characteristic pattern of the mass spectrum~\cite{Barger:2010iv, Ghosh:2010hy}. 
On the other hand, due to the R-parity violation, the usual dark matter (DM) candidate in the MSSM,
  the lightest superpartner (LSP) neutralino is unstable and no longer a DM candidate. 
Thus, we need a new DM candidate to supplement the model.

The U(1)$_{B-L}$ model can be generalized to the U(1)$_X$ model \cite{Appelquist:2002mw},
  where the particle content is the same as the U(1)$_{B-L}$ model,  
  but the U(1)$_X$ charge of a field in the $B-L$ model is extended to be a linear combination of its $B-L$ change ($Q_{B-L}$) and hypercharge ($Q_Y$), namely, $Q_X = x_H \, Q_Y + Q_{B-L}$ with a common parameter $x_H$. 
The U(1)$_{B-L}$ model is defined by the limit of $x_H \to 0$. 
Clearly, the U(1)$_X$ model is anomaly free since both $B-L$ and hypercharge gauge symmetries are 
  free from anomalies in the $B-L$ model. 

In this paper, we propose the minimal SUSY U(1)$_X$ model with R-parity conservation.  
The particle content of the model is the same as the one of the minimal SUSY $B-L$ model discussed above, 
  while we assign an even R-parity to one RHN chiral superfield ($\Psi$)
  and an odd R-parity to the other two RHN chiral superfields. 
The R-parity assignment for the MSSM fields is as usual. 
Because of this parity assignment and the gauge symmetry, the chiral superfield $\Psi$ 
  has no Dirac Yukawa coupling with the lepton doublet fields.  
We consider the case that the $B-L$ symmetry breaking is driven by a VEV of the scalar component of $\Psi$.  
Phenomenological consequences in this model are very different from those of the conventional minimal SUSY $B-L$ model. 
As usual in the MSSM, R-parity is conserved, and hence the LSP neutralino is a candidate for the dark matter (DM). 
In addition to the neutralino in the MSSM, a new DM candidate arises in the model, 
  namely, a linear combination of the fermion component of $\Psi$ and the $B-L$ gaugino,
We will investigate this DM phenomenology and find a parameter region that reproduces the observed DM relic density.   
In the absence of the U(1)$_X$ Higgs superfields carrying U(1)$_X$ charge $\pm2$, 
 no Majorana mass term is generated for RHNs relevant to the mass generation in the SM neutrino sector, 
 and as a result, the SM neutrinos are Dirac particles. 
With only the two RHNs involved in the Dirac Yukawa couplings, 
  the neutrino mass matrix leads to three mass eigenstates:  
  one massless chiral neutrino and two Dirac neutrinos. 
The 2-by-3 neutrino mass matrix includes a sufficient number of free parameters 
  to reproduce the neutrino oscillation data. 
This Dirac nature of the SM neutrinos is distinctive from the neutrinos in the usual $B-L$ model, 
  where the neutrinos are Majorana particles with a hierarchical mass spectrum through the seesaw mechanism. 
In the early universe, the right-handed components of the two Dirac neutrinos are in thermal equilibrium
  with the SM particles through the U(1)$_X$ gauge interaction, and they potentially contribute to extra neutrino species ($\Delta N_\mathrm{eff}$) at the Big Bang Nucleosynthesis (BBN) era 
  \cite{Heeck:2014zfa, FileviezPerez:2019cyn}. 
Applying the cosmological upper bound on $\Delta N_\mathrm{eff}$, we find a lower bound on the scale 
  of U(1)$_X$ symmetry breaking. 
The U(1)$_X$ gauge boson ($Z^\prime$) has been searched by the Large Hadron Collider (LHC) experiment. 
Using the LHC Run-2 final results, we constrain the model parameter region. 
Combining the DM density constraint, the BBN bound and the $Z^\prime$ boson search results, 
  we identify the allowed parameter region. 
We also note that if the $Z^\prime$ boson is discovered at the LHC, 
  the number of left-handed plus right-handed neutrinos can be determined 
  by measuring the invisible decay width of $Z^\prime$ boson.

This paper is organized as follows:
In the next section, we first define our minimal SUSY U(1)$_X$ model with R-parity conservation. 
Then, we introduce the superpotential and soft SUSY breaking terms relevant to our discussion. 
In Sec.~3, we consider the renormalization group (RG) equations for the soft SUSY breaking parameters
  and examine their RG evolution from high energy to low energy. 
We show that the U(1)$_X$ symmetry is radiatively broken by the RG evolution while R-parity remains manifest. 
With a suitable choice of the soft SUSY breaking parameters, the LSP neutralino is identified
  as a linear combination of the fermion component of $\Psi$ and the U(1)$_X$ gaugino. 
In Sec.~4, we investigate this DM physics and find the parameter region that reproduces the observed DM relic density. 
The search for $Z^\prime$ boson resonance at the LHC is discussed in Sec.~5. 
Using the LHC Run-2 final results reported by the ATLAS and CMS Collaborations, 
  we find the LHC constraints on $Z^\prime$ boson mass, U(1)$_X$ gauge coupling and $x_H$. 
We discuss the BBN bound in Sec.~6. 
In Sec.~7, we present the results by combining the cosmological and collider constraints obtained in Secs.~4, 5 and 6. 
We discuss in Sec.~8 how to probe the number of left-handed and right-handed neutrinos in the model 
  by collider experiments. 
The last section is devoted to conclusions and discussions.

%%%%%%%%%%%%%%%%%%%%%%%%%%%%%%
%\vspace{10mm}
\section{Minimal SUSY U(1)$_X$ model with conserved R-parity}
%%%%%%%%%%%%%%%%%%%%%%%%%%%%%%

%%%%%%%%%%%%%%%%%%%%%%%%%%%%%%%%%%%%%%%%%%%%%%%
\begin{table}[t]
	\begin{center}
	\begin{tabular}{|c|ccc|rcr|c|}
 \hline
				& SU(3)$_c$ & SU(2)$_L$ & U(1)$_Y$ & \multicolumn{3}{c|}{U(1)$_X$} & $R$-parity \\
	\hline
	&&&&&&&\\[-12pt]
	$Q_i$    & {\bf 3}   & {\bf 2}& $+1/6$ & $x_Q$ 	& = & $+ \; \frac{1}{3}x_H + \frac{1}{3}x_\Psi$ &$-$ \\[2pt] 
	$U_i^c$    & {\bf 3}$^\ast$ & {\bf 1}& $-2/3$ & $x_U^c$ 		& = & $- \; \frac{4}{3}x_H - \frac{1}{3}x_\Psi$  &$-$ \\[2pt] 
	$D_i^c$    & {\bf 3}$^\ast$ & {\bf 1}& $+1/3$ & $x_D^c$ 		& = & $+ \; \frac{2}{3}x_H - \frac{1}{3}x_\Psi$  &$-$ \\[2pt] 
	\hline
	&&&&&&&\\[-12pt]
	$L_i$    & {\bf 1} & {\bf 2}& $-1/2$ & $x_L$ 	& = & $- \; x_H \;\; - x_\Psi$  &$-$  \\[2pt] 
	$N_{1,2}^c$   & {\bf 1} & {\bf 1}& $0$   & $x_N^c$ 	& = & $+ \; x_\Psi$  &$-$\\[2pt] 
	$\Psi$   & {\bf 1} & {\bf 1}& $0$   & $x_\Psi$ 	& = & $+ \; x_\Psi$  & $+$ \\[2pt] 
	$E_i^c$   & {\bf 1} & {\bf 1}& $+1$   & $x_E^c$ 		& = & $+ \; 2x_H \; + x_\Psi$  &$-$ \\[2pt] 
	\hline
	&&&&&&&\\[-12pt]
	$H_u$      & {\bf 1} & {\bf 2}& $+1/2$  &  $x_{H_u}$ 	& = & $+ \; x_H$\hspace*{12.5mm}  &$+$ \\ 
	$H_d$      & {\bf 1} & {\bf 2}& $-1/2$  &  $x_{H_d}$ 	& = & $- \; x_H$\hspace*{12.5mm}  &$+$ \\ 
 \hline
	\end{tabular}
	\end{center}
	\caption{
	Particle contents of the minimal SUSY U(1)$_X$ model with conserved R-parity.
	$i=1,2,3$ denotes the generation index.
	In addition to the MSSM particle contents, three MSSM singlet superfields ($N_{1,2}^c$ and $\Psi$)
	are introduced.
    We consider $x_\Psi \neq 0$, and set $x_\Psi=1$ without loss of generality. 
	}
	\label{Tab:particle_contents}
	\end{table}
%%%%%%%%%%%%%%%%%%%%%%%%%%%%%%%%%%%%%%%%%%%%%%%

%%%%%%%%%%%%%%%%%%%%%%%%%%
%%%%%%%%%%%%%%%%%%%%%%%%%%

The minimal SUSY U(1)$_X$ model is based on the gauge group of  
  SU(3)$_c\times$SU(2)$_L \times $U(1)$_Y \times$U(1)$_X$.  
In addition to the MSSM particle content, we introduce 
  three chiral superfields, which are singlets under the SM gauge groups but have a U(1)$_X$ charge, $x_\Psi$.
%The new fields are identified as the right-handed neutrino chiral superfields, 
%  and their existence is essential to make the model free from all gauge and gravitational anomalies.   
Their existence is essential to make the model free from all gauge and mixed gauge-gravitational anomalies.   
Unlike direct supersymmetrization of the minimal U(1)$_X$ model, 
   no U(1)$_X$ Higgs superfields carrying U(1)$_X$ charge $\pm2 x_\Psi$ are included in the model. 
The particle content is listed in Table~\ref{Tab:particle_contents}. 
The U(1)$_X$ charge of an MSSM field is defined as a linear combination of its hypercharge and $B-L$ charge 
  such as $Q_X=x_H \, Q_Y + x_\Psi \, Q_{B-L}$. 
For $x_\Psi \neq 0$, we can set $x_\Psi=1$ without loss of generality.\footnote{
We do not consider $x_\Psi =0$. In this case, the phenomenology of the model
 is very different from what we discuss in this paper. 
} 
The key to our proposal is that we assign an even R-parity to $\Psi$ 
  while an odd parity to $N_{1,2}^c$, 
  so that $N_{1,2}^c$ are identified with the RHN superfields relevant to the neutrino mass generation. 
This is in contrast with the minimal SUSY $B-L$ model proposed in Ref.~\cite{Barger:2008wn},
  where all RHN chiral superfields are R-parity odd states and their scalar components develop VEVs 
  by which the U(1)$_{B-L}$ symmetry and R-party at the same time are broken.  
See Ref.~\cite{Okada:2016tzi} for the $B-L$ version of our model, which corresponds to the limit of $x_H \to 0$.

The gauge and R-parity invariant superpotential, which is added to the MSSM one, is 
   only the neutrino Dirac Yukawa coupling, 
\begin{eqnarray}
	{\cal L}_{\rm Yukawa} 
		&=& \sum_{i=1}^{2} \sum_{j=1}^{3} Y_N^{ij} N_i^c H_u L_j .
	\label{Eq:L_Yukawa}
\end{eqnarray}	
Note that R-parity forbids $\Psi$ to have a direct coupling with the MSSM fields.  
After the electroweak symmetry breaking, the neutrino Dirac mass matrix is generated. 
Since this is a 2-by-3 matrix, we have one massless neutrino and two Dirac neutrinos in the model. 
This mass matrix has a sufficient number of free parameters to reproduce the neutrino oscillation data.

Next, we introduce soft SUSY breaking terms for the fields in the U(1)$_X$ sector: 
\begin{eqnarray}
	{\cal L}_{\rm soft} 
		&=& - \left( \frac{1}{2} M_X \lambda_X \lambda_X + {\rm h.c.} \right) 
			- \left( \sum_{i=1}^{2} m_{\tilde{n}_i^c}^2 |\tilde{n}_i^c|^2 + m_\phi^2 |\phi|^2 \right) ,
\end{eqnarray}	
where $\lambda_X$ is the U(1)$_X$ gaugino, and $\tilde{n}_i^c$ and $\phi$ are the scalar components 
  of $N_i^c$ and $\Psi$, respectively.
Since the Dirac Yukawa couplings are very small to reproduce the neutrino oscillation data, 
  we omit terms relevant to the couplings. 
In the next section, we analyze the renormalization group (RG) evolution of the soft SUSY-breaking masses.  
We find that $m_\phi^2$ is driven to be negative while $m_{\tilde{n}_i^c}^2$ remain positive 
  at low energy, and as result only $\phi$ develops non-zero VEV.  
Thus, the U(1)$_X$ symmetry is radiatively broken, but R-parity is conserved at the vacuum.

%%%%%%%%%%%%%%%%%%%%%%%%%%%%%%
%\vspace{10mm}
\section{Radiative U(1)$_X$ symmetry breaking}
%%%%%%%%%%%%%%%%%%%%%%%%%%%%%%

%%%%%%%%%%%%%%%%%%%%%
%%%%%%%%%%%%%%%%%%%%%

%%%%%%%%%%%%%%%%%%%%%%%%%%%%%%%%%%%%%%%%%%%%%%%%
\begin{figure}[t]
\begin{center}
{\includegraphics[scale=0.8]{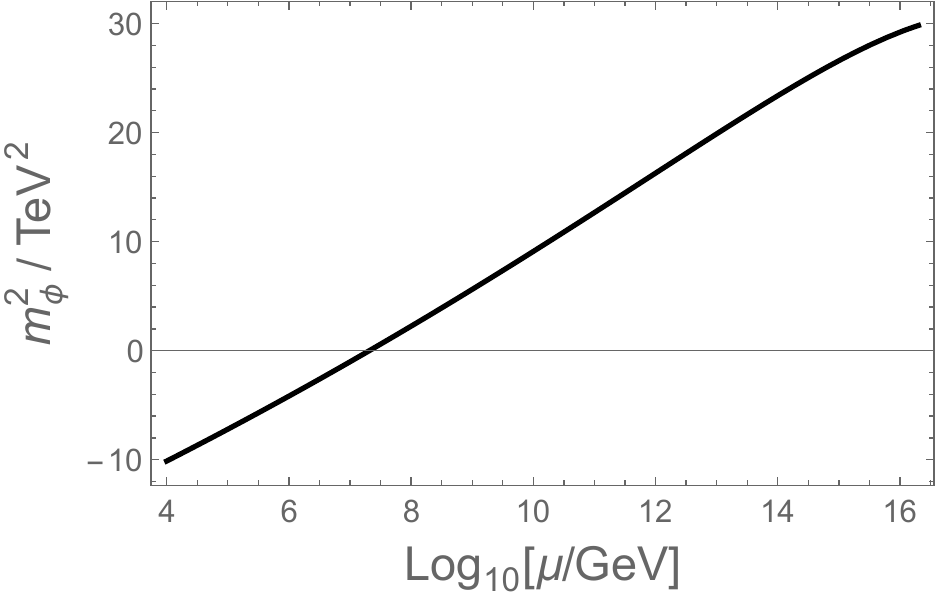}}
\caption{
The RG evolution of the soft SUSY breaking mass $m_\phi^2$ from $M_U$ to low energies. 
}
\label{fig:RG_mphi}
\end{center}
\end{figure}
%%%%%%%%%%%%%%%%%%%%%%%%%%%%%%%%%%%%%%%%%%%%%%%%% 

It is well-known that the electroweak symmetry breaking in the MSSM is triggered by radiative corrections
  which drive the mass squared of the up-type Higgs doublet negative. 
In this mechanism, the electroweak symmetry breaking scale is controlled by the soft SUSY breaking mass scale. 
Thus, the SUSY breaking scale at the TeV naturally results in the electroweak scale of ${\cal O}$(100 GeV). 
In this section, we show that similarly to the MSSM, a radiative U(1)$_X$ symmetry breaking 
  occurs by the RG evolution of the soft SUSY breaking parameter $m_\phi^2$ from high energy to low energy.  
However, note that the mechanism that drives $m_\phi^2$ negative differs from the one in the MSSM 
  where the large top Yukawa coupling plays a crucial role.

For simplicity, we consider the RG equations only for the U(1)$_X$ sector.\footnote{
See Refs.~\cite{Ambroso:2010pe, Ovrut:2015uea} for more elaborate analysis and parameter scans 
  to identify parameter regions.}
At the one-loop level, the RG equations relevant to our discussion are given by
\begin{eqnarray}
    \label{U1x-RGE1}
	\mu \frac{d m_\phi^2}{d \mu} 
		&=& \frac{g_X^2}{16 \pi^2} \left[ 2 \left( \sum_{j=1}^{2} m_{\tilde{n}_j^c}^2 + m_\phi^2 \right) - 8 M_X^2 \right] ,\\
    \label{U1x-RGE2}
	\mu \frac{d m_{\tilde{n}_i^c}^2}{d \mu} 
		&=& \frac{g_X^2}{16 \pi^2} \left[ 2 \left( \sum_{j=1}^{2} m_{\tilde{n}_j^c}^2 + m_\phi^2 \right) - 8 M_X^2 \right] ,\\
    \label{U1x-RGE3}
	\mu \frac{d M_X}{d \mu} 
		&=& \frac{g_X^2}{16 \pi^2} \, 32 M_X ,
\end{eqnarray}	
where the U(1)$_X$ gauge coupling ($g_X$) obeys 
\begin{eqnarray}
	\mu \frac{d g_X}{d \mu} 
		&=& \frac{1}{16 \pi^2} \, 16 g_X^3 .
\end{eqnarray}
Here, we have neglected the contributions from neutrino Dirac Yukawa couplings which are very small.  
The first term inside the bracket on the right-hand side of Eq.~(\ref{U1x-RGE1}), 
   which originates from the U(1)$_X$ $D$-term interaction, 
   plays an essential role in driving $m_\phi^2$ negative at low energy.  
Since squarks and sleptons have U(1)$_X$ charges, their soft squared masses also appear 
 in the RG equations through their $D$-term interactions. 
However, for simplicity, we have omitted them by assuming their squared masses are much smaller
  than $m_{\tilde{n}^c_i}^2$.

To illustrate the radiative U(1)$_X$ symmetry breaking, 
  we numerically solve the RG equations from $M_U=2 \times 10^{16}$ GeV to low energy
  with the following boundary conditions:  
\begin{eqnarray}
 \alpha_{g_X}= 0.0189, \; \;  M_X=16.0 \; {\rm TeV}, \; \;
   m_{\tilde{n}^c_1} = m_{\tilde{n}^c_2} = 25.0 \; {\rm TeV}, \; \; 
   m_{\phi} = 5.46 \; {\rm TeV}, 
\label{BC}
\end{eqnarray}
where $\alpha_{g_X}= \frac{g_X^2}{4 \pi}$. 
Although we do not consider the grand unification of our model in this paper, 
  we took a typical scale of the gauge coupling unification in the MSSM, $M_U =2 \times 10^{16}$ GeV,
  as a reference scale at which the boundary conditions are imposed. 
We show in Fig.~\ref{fig:RG_mphi} the RG evolution of $m_\phi^2$. 
We can see that $m_\phi^2$ becomes negative at low energies, while we find the other squared masses remain positive. 
To arrive at this result, the hierarchy $m_{\tilde{n}^c_i}^2 \gg m_\phi^2, M_X^2$ is crucial, 
  which drives only $m_\phi^2$ negative.

After solving the RG equations, we evaluate the soft SUSY breaking parameters at the U(1)$_X$ symmetry breaking scale, 
  $\mu=v_X=\sqrt{2} \langle \phi \rangle$. 
For $v_X=14$ TeV as a reference, we find 
\begin{eqnarray}
 \alpha_{g_X}= 0.008, \; \;  M_X=6.75 \; {\rm TeV}, \; \; 
   m_{\tilde{n}^c_1} = m_{\tilde{n}^c_2} = 24.2 \; {\rm TeV}, \; \; 
   |m_{\phi}| = 3.18 \; {\rm TeV}.  
\label{LEmass}
\end{eqnarray}
Solving the stationary conditions for the scalar potential, 
\begin{eqnarray}
  V = m_{\tilde{n}^c_1}^2 |\tilde{n}^c_1|^2   +  m_{\tilde{n}^c_2}^2 |\tilde{n}^c_2|^2 + m_{\phi}^2 |\phi|^2
       + \frac{g_X^2}{2} 
       \left(  |\tilde{n}^c_1|^2  + |\tilde{n}^c_2|^2  + |\phi|^2  \right)^2,
\end{eqnarray}
with the above inputs, we find in units of TeV 
\begin{eqnarray}
 \langle \tilde{n}^c_1 \rangle = \langle \tilde{n}^c_2 \rangle =0, \; \; 
 v_X= \sqrt{2} \langle \phi \rangle = \sqrt{2} \frac{|m_\phi(\mu=v_X)|}{g_X(\mu=v_X)}\simeq 14.
\end{eqnarray} 
The last equation indicates the consistency of our analysis. 
For a given choice of the input parameters of $g_X$, $M_X$ and $m_\phi$, 
  we can arrange $m_{\tilde{n}^c_{1,2}}$ to make our analysis consistent. 
We now see that the U(1)$_X$ gauge symmetry is radiatively broken by the VEV of $\phi$. 
Since $\phi$ is R-parity even, R-parity is manifest at the vacuum. 

%In our parameter choice, the $Z^\prime$ boson mass is given by 
%\begin{eqnarray}  
%   m_{Z^\prime} = g_X v_X = \sqrt{2} g_X \langle \phi \rangle. % = 3.5 \; {\rm TeV}. 
%\end{eqnarray}  

%%%%%%%%%%%%%%%%%%%%%%%%%%%%%%
%\vspace{10mm}
\section{New Neutralino Dark Matter}
%%%%%%%%%%%%%%%%%%%%%%%%%%%%%%

%%%%%%%%%%%%%%%%%%%%%%%%%%%%%
%%%%%%%%%%%%%%%%%%%%%%%%%%%%%

Since R-parity is manifest at the vacuum as usual in our model, the LSP neutralino is stable and
  play the DM role. 
In addition to the MSSM neutralinos, a new dark matter candidate arises in our model, 
  namely, a linear combination of the fermion component of $\Psi$ ($\psi$) and the U(1)$_X$ gaugino. 
In this section, assuming this particle is the LSP, we investigate the DM phsyics. 

A scenario of the parity-odd right-handed Majorana neutrino dark matter was first proposed in Ref.~\cite{Okada:2010wd} 
  in the context of the non-SUSY minimal $B-L$ model, 
  where a $Z_2$ parity is introduced and an odd parity is assigned to one right-handed neutrino 
  while the other fields are all parity-even. 
No new particle is introduced other than the particle content in the minimal $B-L$ model. 
Thanks to the conservation of $Z_2$-parity, the parity-odd right-handed neutrino is stable, 
  and hence the DM candidate in the model. 
The phenomenology of this DM particle along with the LHC physics and astrophysics has been investigated 
  in Refs.~\cite{Okada:2010wd, Okada:2012sg, Basak:2013cga,Okada:2016gsh}  
  (see Ref.~\cite{Okada:2018ktp} for a concise review).  
For the U(1)$_X$ extension of the scenario, see, for example, 
  Refs.~\cite{Okada:2016tci, Oda:2017kwl, Okada:2017dqs,Okada:2020cue}. 
A SUSY version of the minimal $B-L$ model with the right-handed neutrino dark matter 
  has been proposed in \cite{Burell:2011wh}. 
The extension of this SUSY model to a U(1)$_X$ model is discussed in Ref.~\cite{Okada:2022stw}.

Although the dark matter scenario in this section shares similar properties 
   with the scenario discussed in Refs.~\cite{Burell:2011wh, Okada:2022stw}, 
   there is a crucial difference, namely, $\psi$ has no Majorana mass on its own, 
   but the DM particle acquires a Majorana mass through the mixing with the U(1)$_X$ gaugino ($\lambda_X$). 
With the U(1)$_X$ symmetry breaking by $\langle \phi \rangle =v_X/\sqrt{2}$, 
   the mass matrix for $\psi$ and $\lambda_X$ is given by
\begin{eqnarray}
	M_\chi &=& \begin{pmatrix} 0 & m_{Z^\prime} \\ m_{Z^\prime} & M_X \end{pmatrix} ,
\end{eqnarray}
which is diagonalized as 
\begin{eqnarray}
	\begin{pmatrix} \chi_\ell \\ \chi_h \end{pmatrix} 
		&=& \begin{pmatrix} \cos \theta & - \sin \theta \\ \sin \theta & \cos \theta \end{pmatrix} 
			\begin{pmatrix} \psi \\ \lambda_X \end{pmatrix} ,
\end{eqnarray}
with $\tan 2 \theta= 2 m_{Z^\prime}/M_X$. 
Here, $m_{Z^\prime} = g_X v_X$ is the $Z^\prime$ boson mass. 
Let us assume that the lighter mass eigenstate ($\chi_\ell$) is the LSP neutralino. 
Since $\psi$ and $\lambda_X$ are the MSSM gauge singlets, 
   possible annihilation processes of a pair of DM particles are very limited. 
Furthermore, given a small U(1)$_X$ gauge coupling and the Majorana nature of the dark matter particle, 
   the annihilation processes via sfermion exchanges are not efficient. 
We find that the most important annihilation process for a pair of DM particles is through 
   the $Z^\prime$ boson resonance in the $s$-channel with the DM mass being close to half of the $Z^\prime$ boson mass. 
To realize this situation, we set $M_X \simeq (3/2) m_{Z^\prime}$, so that the lightest mass eigenvalue is 
  given by $m_{DM} \simeq m_{Z^\prime}/2$ and $\cos^2 \theta \simeq 0.8$. 
Our parameter choice in the previous section is suitable for this setup, 
  $M_X= (3/2) m_{Z^\prime}=6.75$ TeV for $m_{Z^\prime}=4.5$ TeV.

We now calculate the DM relic density by integrating out the Boltzmann equation,
\begin{eqnarray}
  \frac{dY}{dx} &=& -\frac{xs \langle \sigma v \rangle}{H(m_{\rm DM})} (Y^2 - Y_{\rm EQ}^2),
  \label{Eq:Boltzmann}
\end{eqnarray}
where the temperature of the universe is normalized by the DM mass $x=m_{DM}/T$, 
   $H(m_{DM})$ is the Hubble parameter at $T=m_{DM}$, 
   $Y$ is the yield (the ratio of the dark matter number density to the entropy density $s$) of 
   the DM particle, $Y_{EQ}$ is the yield of the DM particle in thermal equilibrium, 
  and $\langle \sigma v \rangle$ is the thermal average of the dark matter annihilation cross section times relative velocity. 
Explicit formulas of the quantities involved in the Boltzmann equation are as follows: 
\bea 
s &=& \frac{2  \pi^2}{45} g_\star \frac{m_{DM}^3}{x^3} ,  \nonumber \\
H(m_{DM}) &=&  \sqrt{\frac{\pi^2}{90} g_\star} \frac{m_{DM}^2}{M_{P}},  \nonumber \\ 
s Y_{EQ}&=& \frac{g_{DM}}{2 \pi^2} \frac{m_{DM}^3}{x} K_2(x),   
\eea
where $M_{P}=2.43 \times 10^{18}$  GeV is the reduced Planck mass, 
   $g_{DM}=2$ is the number of degrees of freedom for the Majorana dark matter particle, 
   $g_\star$ is the effective total number of degrees of freedom for particles in thermal equilibrium 
   (in the following analysis, we use $g_\star=106.75$ for the SM particles),  
   and $K_2$ is the modified Bessel function of the second kind.   
In our scenario, a DM pair annihilates into the SM particles 
   dominantly through the $Z^\prime$ boson exchange in the $s$-channel.  
The thermal average of the annihilation cross section is given by 
\begin{eqnarray}
  \langle \sigma v \rangle 
	  &=& (sY_{\rm EQ})^{-2} g_{\rm DM}^2 \frac{m_{\rm DM}}{64 \pi^4 x}
		   \int_{4m_{\rm DM}^2}^{\infty} ds \, 2 (s- 4 m_{DM}^2) \sigma(s)    \sqrt{s}
		   K_1\left( \frac{x \sqrt{s}}{m_{\rm DM}} \right),
\end{eqnarray}
where  $\sigma(s)$ is the total annihilation cross section, and $K_1$ is the modified Bessel function of the first kind. 
The total cross section of the dark matter annihilation process $\chi_\ell \chi_\ell \to Z^\prime \to f {\bar f}$ 
  ($f$ denotes the SM fermions plus two right-handed neutrinos) 
   is calculated as 
\begin{eqnarray}
\sigma(s) &=& \frac{\pi}{3} \alpha_{g_X}^2 \cos^4 \theta
	\frac{\sqrt{s (s-4m_{\rm DM}^2)}}{(s-m_{Z^\prime}^2)^2+m_{Z^\prime}^2 \Gamma_{Z^\prime}^2} \nonumber \\
	&\times& \left[ \frac{103x_H^2 + 86x_H + 37}{3} 
		+ \frac{17x_H^2 + 10x_H + 2 + (7x_H^2 + 20x_H + 4)\frac{m_t^2}{s}}{3} 
	 		\sqrt{1-\frac{4m_t^2}{s}} \right.  \nonumber \\
	&& \hspace{20mm}
		\left. + 18 x_H^2 \frac{(s-m_{Z^\prime}^2)^2}{s (s-4m_{\rm DM}^2)}
		\frac{m_{\rm DM}^2 m_t^2}{m_{Z^\prime}^4}
		\sqrt{1-\frac{4m_t^2}{s}} \right],
\end{eqnarray}
where all final state fermion masses have been neglected except for the top quark mass $m_t$,
   and we have assumed that all MSSM particles are heavier than the DM particle. 
The total decay width of $Z^\prime$ boson is given by 
\begin{eqnarray}
\Gamma_{Z^\prime} 
%    &=& 
%	\frac{\alpha_{g_X} m_{Z^\prime}}{6} \!\!
%	\left[ \frac{103x_H^2 + 86x_H + 37}{3} 
%		+ \frac{17x_H^2 + 10x_H + 2 + (7x_H^2 + 20x_H + 4)\frac{m_t^2}{m_{Z^\prime}^2}}{3} 
%	 		\sqrt{1-\frac{4m_t^2}{m_{Z^\prime}^2}} \right.  \nonumber \\
%	&& \hspace{15mm}
%		\left. + 2 \left( 1 - \frac{4m_N^2}{m_{Z^\prime}^2} \right)^{\frac{3}{2}} 
%			\theta \left( \frac{m_{Z^\prime}^2}{m_N^2} -4\right) 
%				+  \cos^4 \theta \left( 1 - \frac{4m_{\rm DM}^2}{m_{Z^\prime}^2} \right)^{\frac{3}{2}} 
%			\theta \left( \frac{m_{Z^\prime}^2}{m_{\rm DM}^2} -4\right) \right.  \nonumber \\
%	&& \hspace{80mm}
%		\left. + 2 x_H^2 \sqrt{\lambda[a, b ,c]} 
%                \left(  \lambda[a, b, c] + 12 b  \right) \right] \nonumber \\
%    \nonumber \\
    &=& 
	\frac{\alpha_{g_X} m_{Z^\prime}}{6} \!\!
	\left[ \frac{103x_H^2 + 86x_H + 43}{3} 
		+ \frac{17x_H^2 + 10x_H + 2 + (7x_H^2 + 20x_H + 4)\frac{m_t^2}{m_{Z^\prime}^2}}{3} 
	 		\sqrt{1-\frac{4m_t^2}{m_{Z^\prime}^2}} \right.  \nonumber \\
	&& \hspace{15mm}
		\left. +  \cos^4 \theta \left( 1 - \frac{4m_{\rm DM}^2}{m_{Z^\prime}^2} \right)^{\frac{3}{2}} 
			\Theta \left( \frac{m_{Z^\prime}^2}{m_{\rm DM}^2} -4\right) 
            + 2 x_H^2 \sqrt{\lambda[a, b ,c]} 
                \left(  \lambda[a, b, c] + 12 b  \right) \right]. 
\label{Eq:DecayWidthZp}
\end{eqnarray}
where $\Theta$ is the Heaviside step function, 
 and the last term in the bracket on the right-hand side corresponds to $Z^\prime \to Z \, h$ with 
 $\lambda[x, y, z]= x^2+y^2+z^2-2xy-2yz-2zx$, 
  $a=1$, $b= \left( \frac{m_Z}{m_{Z^\prime}} \right)^2 \ll1 $ 
  and $c= \left( \frac{m_h}{m_{Z^\prime}} \right)^2 \ll1 $~\cite{Das:2020rsr}, respectively, 
  with the $Z$ boson mass $m_Z=91.2$ GeV and the Higgs boson mass $m_h=125$ GeV.

%%%%%%%%%%%%%%%%%%%%%%%%%%%%%%%%%%%%%%%%%%%%%%%%
%   Fig DM abundance 
%%%%%%%%%%%%%%%%%%%%%%%%%%%%%%%%%%%%%%%%%%%%%%%%
\begin{figure}[t]
\begin{center}
{\includegraphics[scale=1.0]{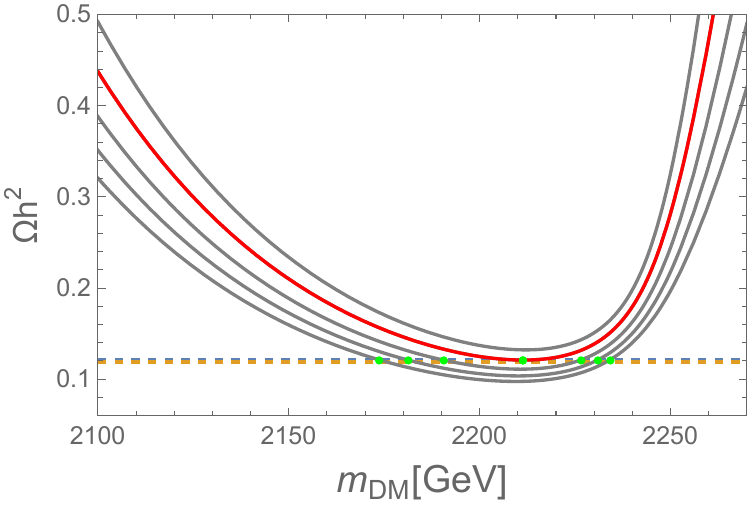}}
\caption{
The relic abundance of the dark matter particle  
  as a function of the dark matter mass ($m_{DM}$)
  for $x_H=-0.5$, $m_{Z^\prime}=4.5$ TeV, $\cos^2 \theta=0.8$ 
  and various values of the gauge coupling,  
  $\alpha_{g_X}=0.007$, $0.00793$, $0.009$, $0.010$ and $0.011$ 
  (solid lines from top to bottom). 
The region between two horizontal dashed lines denote the range of the observed dark matter relic abundance, 
  $0.119 \leq \Omega_{DM} h^2 \leq 0.121$ (68\% limit) \cite{Planck:2018vyg}. 
}
\label{Fig:relic}
\end{center}
\end{figure}
%%%%%%%%%%%%%%%%%%%%%%%%%%%%%%%%%%%%%%%%%%%%%%%%%

We solve the Boltzmann equation numerically to find the asymptotic value of the yield $Y(\infty)$. 
The dark matter relic abundance is evaluated as 
\bea 
  \Omega_{DM} h^2 =\frac{m_{DM} s_0 Y(\infty)} {\rho_c/h^2}, 
\eea 
  where $s_0 = 2890$ cm$^{-3}$ is the entropy density of the present universe, 
  and $\rho_c/h^2 =1.05 \times 10^{-5}$ GeV/cm$^3$ is the critical density.
In our analysis, only three parameters, 
   namely $\alpha_{g_X}$, $m_{Z^\prime}$ and $m_{DM}$, are involved. 
Note that the mixing angle $\theta$ is not a free parameter since it is determined
   once $m_{Z^\prime}$ and $m_{DM}$ are fixed.   
We identify the parameter region so as to reproduce the observed DM relic density of $\Omega_{DM} h^2 =0.12$ from the Planck results~\cite{Planck:2018vyg}.

As mentioned above, a sufficiently large annihilation cross section is achieved 
   only if $m_{DM} \simeq m_{Z^\prime}/2$. 
Thus, we focus on the DM mass in this region and in this case $\cos^2 \theta \simeq 0.8$. 
For $x_H=-0.5$, $m_{Z^\prime}=4.5$ TeV and $\cos^2 \theta=0.8$, 
  Fig.~\ref{Fig:relic} shows the resultant DM relic density
  as a function of its mass $m_{DM}$ 
  for various values of the gauge coupling, $\alpha_{g_X}=0.007$, $0.00793$, $0.009$, $0.010$ and $0.011$.
The region between two horizontal dashed lines is the allowed region, 
  $0.119 \leq \Omega_{DM} h^2 \leq 0.121$ (68\% limit), 
  from the Planck satellite experiment~\cite{Planck:2018vyg}. 
For each contour, the DM masses reproducing the observed relic abundance are marked as green dots.     
We have confirmed that the observed relic abundance can be reproduced 
  when the dark matter mass is close to half of the $Z^\prime$ boson mass. 
Note that we find the lower bound on $\alpha_{g_X}$ once the values of $x_H$ and $m_{Z^\prime}$ are fixed. 
In Fig.~\ref{Fig:relic}, we can see $\alpha_{g_X} \leq  0.00793$.

%%%%%%%%%%%%%%%%%%%%%%%%%%%%%%
%\vspace{10mm}
\section{LHC Run-2 constraints}
%%%%%%%%%%%%%%%%%%%%%%%%%%%%%%

%%%%%%%%%%%%%%%%%%%   Sequential SM Z' boson mass by LHC 2019  %%%%%%%%%%%%%%%%%%%%
%\begin{figure}[t]
%%%%%%%%%%%%%   ATLAS 13 TeV 2019 result  %%%%%%%%%%%%%%%%%%%%%%%%%%
%\begin{minipage}{0.5\linewidth}
%\begin{center}
%\includegraphics[width=0.95\linewidth]{CMS2019.pdf}
%\subcaption{}\label{Fig:LHC2019_CMS}
%\end{center}
%\end{minipage}
%%%%%%%%%%%%%   CMS 13 TeV 2019 result  %%%%%%%%%%%%%%%%%%%%%%%%%%
%\begin{minipage}{0.5\linewidth}
%\begin{center}
%\includegraphics[width=0.95\linewidth]{ATLAS2019.pdf}
%\subcaption{}\label{Fig:LHC2019_ATLAS}
%\end{center}
%\end{minipage}
%\caption
%{
%\subref{Fig:LHC2019_CMS} The cross section ratio as a function of the $Z^\prime_{SSM}$ mass (solid line) 
%     with the $k$-factor function of $m_{Z^\prime_{SSM}}$, which is completely covered the CMS 2019 result (dotted line)
%     from the combined dielectron and dimuon channels in Ref.~\cite{CMS:2019tbu}. 
%\subref{Fig:LHC2019_ATLAS} The cross section as a function of $m_{Z^\prime_{SSM}}$ (solid line) 
%     with the $k$-factor function of $m_{Z^\prime_{SSM}}$, which is also covered the ATLAS 2019 result (dotted line)
%     from the combined dielectron and dimuon channels in Ref.~\cite{ATLAS:2019erb}. 
%}
%\label{Fig:LHC2019}
%\end{figure}
%%%%%%%%%%%%%%%%%%%%%%%%%%%%%%%%%%%%%%%%%%%%%%%%%%%%

The ATLAS and the CMS Collaborations have searched for $Z^\prime$ boson resonance 
  at the LHC Run-2 with $\sqrt{s}=13$ TeV. 
The most stringent bounds on the $Z^\prime$ boson production cross section times branching ratio 
  have been obtained by using the dilepton final state. 
For the so-called sequential SM $Z^\prime$  ($Z^\prime_{SSM}$) \cite{Barger:1980ti},
  which has exactly the same couplings with the SM fermions as those of the SM $Z$ boson, 
  we have $m_{Z^\prime_{SSM}} \geq 5.1$ TeV from the ATLAS 2019 results~\cite{ATLAS:2019erb} and 
  $m_{Z^\prime_{SSM}} \geq 5.15$ TeV from the CMS 2019 results~\cite{CMS:2019tbu}, respectively.  
We interpret these ATLAS and CMS results into the U(1)$_X$ $Z^\prime$ boson case  
  and derive the LHC Run-2 constraints on $x_H$, $\alpha_{g_X}$ and $m_{Z^\prime}$.

%%%%%%%%%%%%%%%%%%%%%%%%%%%%%%
\subsection{Bounds from CMS 2019 results}
%%%%%%%%%%%%%%%%%%%%%%%%%%%%%%
We calculate the dilepton production cross section
   for the process $pp \to Z^\prime +X \to \ell^{+} \ell^{-} +X$. 
The differential cross section with respect to the invariant mass $M_{\ell \ell}$ of the final state dilepton 
   is described as
\begin{eqnarray}
\frac{d \sigma}{d M_{\ell \ell}}
	= \sum_{a,b}
		\int^{1}_{\frac{M^2_{\ell \ell}}{E^2_{\rm CM}}} d x_1 \frac{2M_{\ell \ell}}{x_1 E^2_{\rm CM}} 
		f_{a}(x_{1}, M^2_{\ell \ell}) f_{b}\left(\frac{M^2_{\ell \ell}}{x_{1} E^2_{\rm CM}}, M^2_{\ell \ell} \right)  
		\hat{\sigma} (\bar{q} q \to Z^\prime \to  \ell^+ \ell^-),
\label{CrossLHC}
\end{eqnarray}
where $f_a$ is the parton distribution function (PDF) for a parton $a$, 
  and $E_{\rm CM} =13$ TeV is the center-of-mass energy of the LHC Run-2.
In our numerical analysis, we employ CTEQ6L~\cite{Pumplin:2002vw} for the PDFs. 
In the U(1)$_X$ model, the cross sections for the colliding partons are given by 
\begin{eqnarray}
\hat{\sigma} (\bar{u} u \rightarrow Z^\prime \to \ell^+ \ell^-)  
	&=& \frac{\pi \alpha_{g_X}^2}{81} 
		\frac{M_{\ell \ell}^2}{(M_{\ell \ell}^2-m_{Z^\prime}^2)^2 + m_{Z^\prime}^2 \Gamma_{Z^\prime}^2}
		(85x_H^4 + 152x_H^3 + 104x_H^2 + 32x_H + 4), 
\nonumber\\
\hat{\sigma}  (\bar{d} d \rightarrow Z^\prime \to \ell^+ \ell^-) 
	&=& \frac{\pi \alpha_{g_X}^2}{81} 
		\frac{M_{\ell \ell}^2}{(M_{\ell \ell}^2-m_{Z^\prime}^2)^2 + m_{Z^\prime}^2 \Gamma_{Z^\prime}^2}
		(25x_H^4 + 20x_H^3 + 8x_H^2 + 8x_H + 4), 
\label{CrossLHC2}
\end{eqnarray}
where the total decay width of the $Z^\prime$ boson is given in Eq.~(\ref{Eq:DecayWidthZp}).

In interpreting the CMS results into the U(1)$_X$ $Z^\prime$ boson case,  
   we follow the strategy in Refs.~\cite{Okada:2016gsh,Oda:2017kwl}.
In order to take into account the difference between the parton distribution functions
  used in the CMS analysis and our analysis, and QCD corrections of the process,
  we scale our resultant cross section by a $k$-factor as a function of $m_{Z^\prime}$, 
  which is determined so as to fit our cross section calculation for the $Z^\prime_{SSM}$ boson 
  to the theoretical line presented in Ref.~\cite{CMS:2019tbu}.       
According to the analysis by the CMS Collaboration,  
   we integrate the differential cross section for the range of 
  $0.97 \; m_{Z^\prime} \leq  M_{\ell \ell} \leq  1.03 \; m_{Z^\prime}$~\cite{CMS:2019tbu} 
  and obtain the cross section as a function of $x_H$, $\alpha_{g_X}$ and $m_{Z^\prime}$. 
In the CMS analysis, the limits are set on the ratio of the cross sections mediated by $Z'$ and $Z$ bosons, 
\begin{eqnarray}
R_{\sigma} &=&
	\frac{\sigma (pp \to Z^\prime+X \to \ell \ell +X)}
		{\sigma (pp \to Z+X \to \ell \ell +X)},
\end{eqnarray}
 as a function of $m_{Z^\prime}$, where the cross section $\sigma (pp \to Z+X \to \ell \ell +X)$ 
 is evaluated in the mass window of $60$ GeV$\leq  M_{\ell \ell} \leq 120$ GeV 
 which is predicted to be $1928$ pb at the LHC Run-2~\cite{CMS:2015nhc}.

%%%%%%%%%%%%%%%%%%%%%%%%%%%%%%
\subsection{Bounds from ATLAS 2019 results}
%%%%%%%%%%%%%%%%%%%%%%%%%%%%%%
To interpret the ATLAS 2019 results into the U(1)$_X$ model, we use the narrow width approximation
   to evaluate the $Z^\prime$ boson production cross section. 
In this approximation, the total $Z^\prime$ boson production cross section is given by
\begin{eqnarray}
	\sigma (p p \to Z^\prime)
		&=& 2 \sum_{q, \bar{q}} \int d x \int d y \: f_{q}(x, Q) \: f_{\bar{q}} (y, Q)  
			\: \hat{\sigma} (\hat{s}),
	\label{Eq:Zp_CrossSection}
\end{eqnarray}
  where $\hat{s}= x y s$ is the invariant mass squared of the colliding partons (quarks) 
   with a center-of-mass energy $\sqrt{s}=13$ TeV, 
   and the $Z^\prime$ boson production cross section at the parton level is expressed as
\begin{eqnarray}
	\hat{\sigma} (\hat{s})
		&=& \frac{4 \pi^2}{3} \frac{\Gamma (Z^\prime \to q \bar{q})}{m_{Z^\prime}} \: \delta (\hat{s} - m_{Z^\prime}^2) .
	\label{Eq:Zp_CrossSection_atParton}
\end{eqnarray}
Here, $\Gamma(Z^\prime \to q \bar{q})$ is the $Z^\prime$ boson partial decay width to $q \bar{q}$. 
For the up-type and down-type quarks, respectively, 
\begin{eqnarray}
	\Gamma (Z^\prime \to u \bar{u})
		&=& \frac{g_X^2}{72 \pi} ( 2 + 10 x_H + 17 x_H^2) m_{Z^\prime}, \nonumber \\
	\Gamma (Z^\prime \to d \bar{d})
		&=& \frac{g_X^2}{72 \pi} ( 2 - 2 x_H + 5 x_H^2) m_{Z^\prime}.
	\label{Eq:Zp_DecayWidth}
\end{eqnarray}
We then simply express the dilepton production cross section as
\begin{eqnarray}
  \sigma( pp \to Z^\prime \to \ell \bar{\ell}) 
  &\simeq& \sigma( pp \to Z^\prime) \times {\rm BR}(Z^\prime \to \ell \bar{\ell})
\end{eqnarray}
with the branching ratio to $Z^\prime \to \ell \ell$, where $\ell=e$ or $\mu$. 
In order to take into account the difference of the parton distribution functions
  used in the ATLAS analysis and our analysis, and QCD corrections of the process,
  we scale our resultant cross section by a $k$-factor as a function of $m_{Z^\prime}$, 
  which is determined so as to fit our cross section calculation for $Z^\prime_{SSM}$ boson  
  to the theoretical line presented in \cite{ATLAS:2019erb}. 
%Our result for the $Z^\prime_{SSM}$ model is shown as the solid line
%   in Fig.~\ref{Fig:LHC2019}\subref{Fig:LHC2019_ATLAS},  
%   which exactly overlaps with the ATLAS 2019 result (diagonal dotted line) in Ref. \cite{ATLAS:2019erb}.
%The narrow width approximation is valid 
%    only for satisfying the condition that the ratio of the decay width to $Z^\prime$ boson mass is very small 
%    ($\Gamma_{Z^\prime} / m_{Z^\prime} \ll 1$).
%In Fig.~\ref{Fig:LHC2019}\subref{Fig:LHC2019_ATLAS}, 
%    we set $\Gamma_{Z^\prime} / m_{Z^\prime} < 3\%$, 
%    and depict the upper bound (lighter red horizontal solid line) as a reference.

%%%%%%%%%%%%%%%%%%%%%%%%%%%%%%
%\vspace{10mm}
\section{BBN Bound}
%%%%%%%%%%%%%%%%%%%%%%%%%%%%%%
The right-handed neutrinos ($\nu_R$) in our model are in thermal equilibrium with the SM particles 
  in the early universe through $Z^\prime$ boson interaction. 
The thermal averaged cross section at the temperature $T \ll m_{Z^\prime}$ is expressed as
\begin{eqnarray}
	\langle \sigma v \rangle 
		&=& \sum_{f} \langle \sigma v (f \bar{f} \leftrightarrow \nu_R \overline{\nu_R} ) \rangle ,
\end{eqnarray}
  with
\begin{eqnarray}
	\langle \sigma v (u \bar{u} \leftrightarrow \nu_R \overline{\nu_R} ) \rangle
		&\simeq& \frac{g_X^4 N_c T^2}{18 \pi m_{Z^\prime}^4} ( 2 + 10 x_H + 17 x_H^2 ) ,  \\
	\langle \sigma v (d \bar{d} \leftrightarrow \nu_R \overline{\nu_R} ) \rangle
		&\simeq& \frac{g_X^4 N_c T^2}{18 \pi m_{Z^\prime}^4} ( 2 - 2 x_H + 5 x_H^2 ) ,  \\
	\langle \sigma v (\ell \bar{\ell} \leftrightarrow \nu_R \overline{\nu_R} ) \rangle
		&\simeq& \frac{g_X^4 T^2}{2 \pi m_{Z^\prime}^4} ( 2 + 6 x_H + 5 x_H^2 ) ,  \\
	\langle \sigma v (\nu_L \overline{\nu_L} \leftrightarrow \nu_R \overline{\nu_R}) \rangle
		&\simeq& \frac{g_X^4 T^2}{2 \pi m_{Z^\prime}^4} ( 1 + x_H )^2 ,  
\end{eqnarray}
	and the color factor $N_c = 3$ for quarks. 
The decoupling temperature $T_{\rm dec}$ of $\nu_R$ from the thermal bath is evaluated by
\begin{eqnarray}
	\langle \sigma v \rangle n_{\nu_R} |_{T = T_{\rm dec}}
		&=& H(T_{\rm dec}) ,
\end{eqnarray}
where $H = \sqrt{\frac{\rho_{\rm rad}}{3 M_{\rm P}^2}}$, and $\rho_{\rm rad} = \frac{\pi^2}{30}  g_\star T^4$. 

After the decoupling, we parameterize the energy density of $\nu_R$ at the BBN era as
\begin{eqnarray}
\rho_{\nu_R} =  \Delta N_\mathrm{eff} \, \frac{7}{4} \, \frac{\pi^2}{30} \, T_{\rm BBN}^4 ,
\end{eqnarray}
where $T_{\rm BBN} \simeq 1$ MeV is the photon temperature.  
The total effective number of neutrinos is expressed as
\begin{eqnarray}
    N_\mathrm{eff} = N_\mathrm{eff}^\mathrm{SM} + \Delta N_\mathrm{eff}, 
\end{eqnarray}
Using the Planck result \cite{Planck:2018vyg},
\begin{eqnarray}
    N_\mathrm{eff} = 2.99 ^{+0.34}_{-0.33}
\end{eqnarray}
    at $95$\% C.L. ($2.99 \pm 0.17$ at 68\% C.L.), and the SM prediction
    $N_\mathrm{eff}^\mathrm{SM} = 3.046$ 
(see e.g., Refs.~\cite{Escudero:2018mvt,Bennett:2019ewm,Escudero:2020dfa,Akita:2020szl,Bennett:2020zkv}),     
we obtain the upper bound on $\Delta N_\mathrm{eff} \leq 0.284$ at $95$\% C.L. ($0.114$ at $68$\% C.L.).     
The contribution of the two right-handed neutrinos is estimated as  
\begin{eqnarray}
    \Delta N_\mathrm{eff} = 2 \times \left( \frac{g_\star(T_{BBN})}{g_\star(T_{\rm dec})} \right)^{4/3}, 
\end{eqnarray}  
where $g_\star(T_{BBN})=10.75$, and $g_\star(T_{\rm dec})$ is the total relativistic degrees of freedom of 
 SM particles at $T=T_{\rm dec}$. 
To satisfy the upper bound on $\Delta N_\mathrm{eff}$, we find a lower bound on $T_{\rm dec}$,
 which leads to a lower bound on $v_X = m_{Z^\prime}/g_X$ for a fixed $x_H$ value.  
Note that in our model, the minimum value of $\Delta N_\mathrm{eff}$ is found to be $\Delta N_\mathrm{eff} \geq 0.0937$
  by using $g_\star(T_{\rm dec} > m_t) =106.75$. 
This value can be tested in the upcoming the CMB-S4 experiment \cite{CMB-S4:2016ple}. 

%%%%%%%%%%%%%%%%%%%%%%%%%%%%%%%
\section{Combined results}
%%%%%%%%%%%%%%%%%%%%%%%%%%%%%%%
%\label{Sec_allowed_region}
%%%%%%%%%%%%%%%%%%%   m_Z' and alpha_gx  %%%%%%%%%%%%%%%%%%%%
\begin{figure}[t]
\begin{center}
\includegraphics[width=0.8\linewidth]{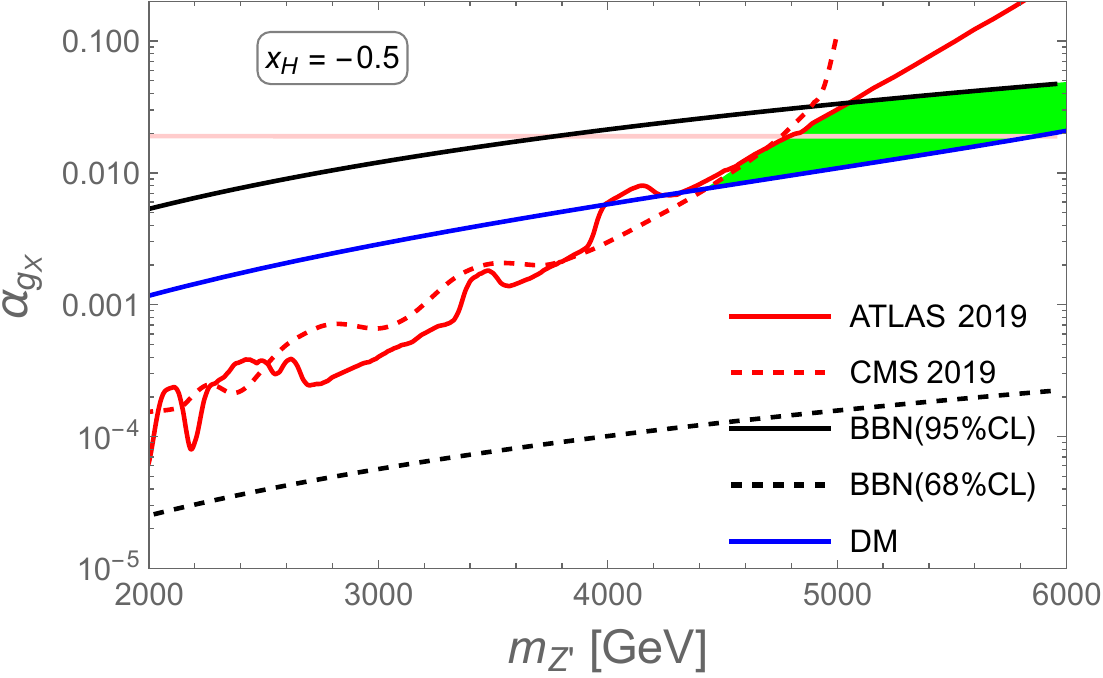}
\caption
{
The solid blue line is the lower bound on $\alpha_{g_{X}}$ as a function of $m_{Z^\prime}$ 
    to reproduce the observed DM relic density of the Planck result \cite{Planck:2018vyg}. 
The solid (dashed) red line shows the upper bound on $\alpha_{g_X}$ 
    obtained by the ATLAS 2019 results~\cite{ATLAS:2019erb}
   (the CMS 2019 results~\cite{CMS:2019tbu}). 
The lighter red horizontal solid line denotes the upper bound 
    for the narrow width approximation to be valid, where we have set $\Gamma_{Z^\prime} / m_{Z^\prime} < 3\%$.
The black solid (dashed) line shows the 95\% C.L. (68\% C.L.) BBN bound on $\Delta N_\mathrm{eff}$. 
The green-shaded region satisfies all constraints.
}
\label{Fig:mZp_alpha}
\end{center}
\end{figure}
%%%%%%%%%%%%%%%%%%%%%%%%%%%%%%%%%%%%%%%%%%%%%%%%%%%%
%%%%%%%%%%%%%%%%%%%%%%%%%%%%%%%%%%%%%%%%%%%%%%%%%%%%
%%%%%%%%%%%%%%%%%%%   x_H and alpha_gx  %%%%%%%%%%%%%%%%%%%%
\begin{figure}[htbp]
%%%%%%%%%%%%%   4 TeV result  %%%%%%%%%%%%%%%%%%%%%%%%%%
\begin{minipage}{0.5\linewidth}
\begin{center}
\includegraphics[width=0.95\linewidth]{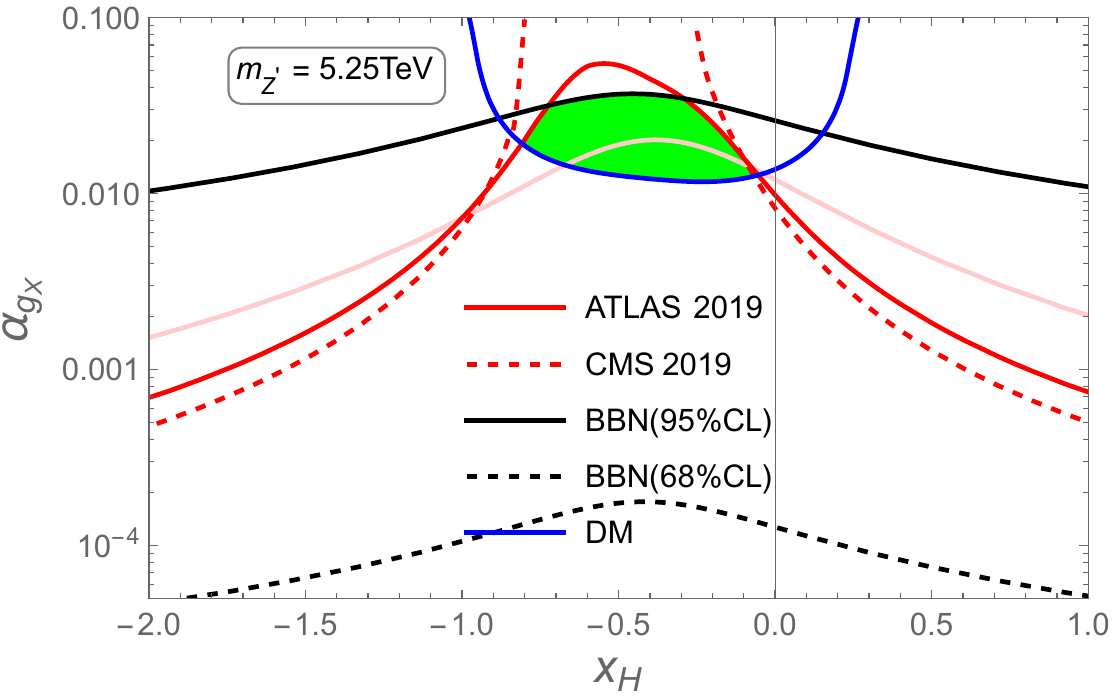}
\subcaption{}\label{Fig:xH_alpha_mzp=5250}
\vspace{5mm}
\end{center}
\end{minipage}
%%%%%%%%%%%%%   3.75 TeV result  %%%%%%%%%%%%%%%%%%%%%%%%%%
\begin{minipage}{0.5\linewidth}
\begin{center}
\includegraphics[width=0.95\linewidth]{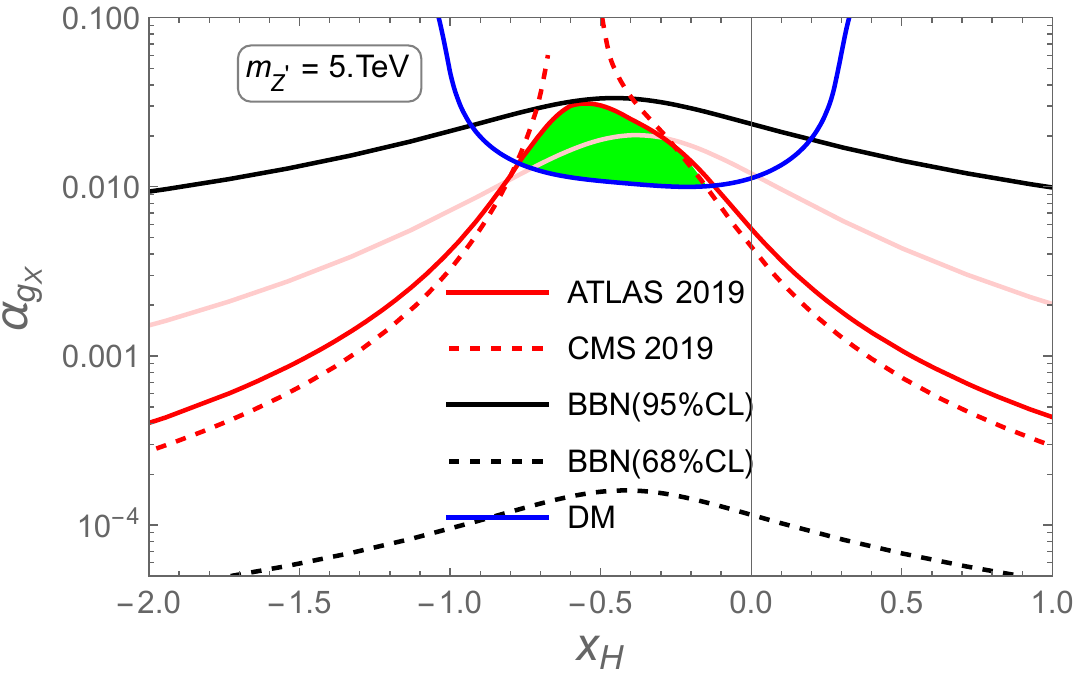}
\subcaption{}\label{Fig:xH_alpha_mzp=5000}
\vspace{5mm}
\end{center}
\end{minipage}
%%%%%%%%%%%%%   3.5 TeV result  %%%%%%%%%%%%%%%%%%%%%%%%%%
\begin{minipage}{0.5\linewidth}
\begin{center}
\includegraphics[width=0.95\linewidth]{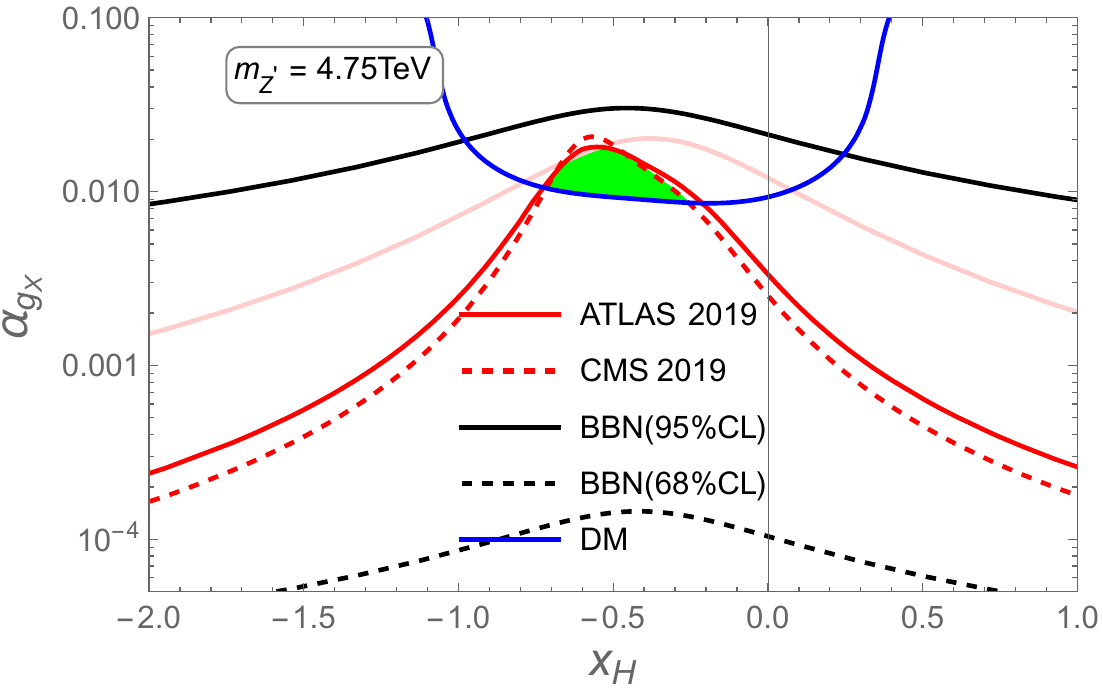}
\subcaption{}\label{Fig:xH_alpha_mzp=4750}
\end{center}
\end{minipage}
%%%%%%%%%%%%%   3.5 TeV result  %%%%%%%%%%%%%%%%%%%%%%%%%%
\begin{minipage}{0.5\linewidth}
\begin{center}
\includegraphics[width=0.95\linewidth]{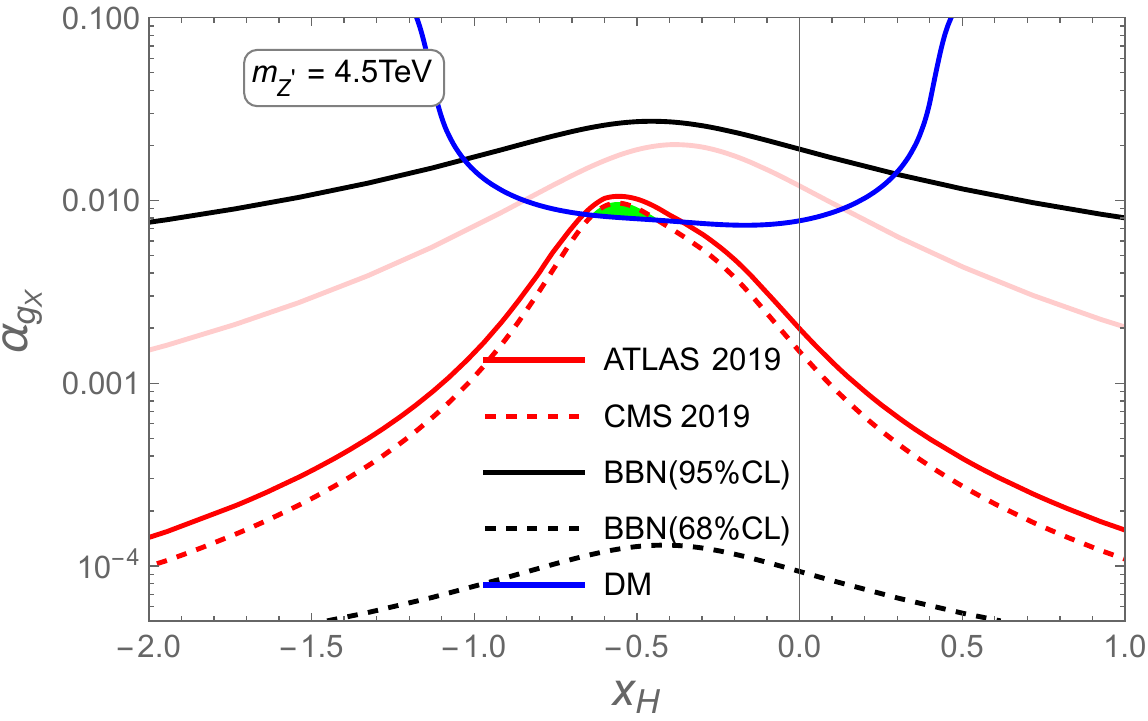}
\subcaption{}\label{Fig:xH_alpha_mzp=4500}
\end{center}
\end{minipage}
\caption{
Allowed parameter regions in the ($x_H$, $\alpha_{g_X}$)-plain 
   for various $m_{Z^\prime}$ values. 
\subref{Fig:xH_alpha_mzp=5250} is for $m_{Z^\prime}=5.25$ TeV. 
The (blue) convex-downward solid line shows the cosmological lower bound on $\alpha_{g_X}$ as a function of $x_H$.
The (red) convex-upward solid (dashed) line shows the upper bound on $\alpha_{g_X}$ 
    obtained from the $Z^\prime$ boson search by the ATLAS~\cite{ATLAS:2019erb}
    (CMS~\cite{CMS:2019tbu}) Collaboration,
    along with the upper bound (lighter red horizontal solid line) 
    for which the narrow width approximation is valid (we have set $\Gamma_{Z^\prime} / m_{Z^\prime} \leq 3\%$).
The black solid (dashed) line shows the 95\% C.L. (68\% C.L.) BBN bound on $\Delta N_\mathrm{eff}$. 
The green shaded region satisfies all constraints.
\subref{Fig:xH_alpha_mzp=5000}, \subref{Fig:xH_alpha_mzp=4750}
   and \subref{Fig:xH_alpha_mzp=4500} are the same as \subref{Fig:xH_alpha_mzp=5250},
   but $m_{Z^\prime}=5$ TeV, 4.75 TeV and 4.5 TeV, respectively.
}
\label{Fig:xH_alpha}
\end{figure}
%%%%%%%%%%%%%%%%%%%%%%%%%%%%%%%%%%%%%%%%%%%%%%%%%%%%
Now we combine all the constraints that we have obtained in the previous sections
   from the DM relic density, the LHC constraints, and the BBN bound. 
In Fig.~\ref{Fig:mZp_alpha}, we show the allowed region in the ($m_{Z^\prime}$, $\alpha_{g_X}$)-plain 
   for fixed $x_H=-0.5$, as an example. 
The solid blue line is the lower bound on $\alpha_{g_{X}}$ as a function of $m_{Z^\prime}$ 
    to reproduce the observed DM relic density of the Planck result \cite{Planck:2018vyg}. 
The solid (dashed) red line shows the upper bound on $\alpha_{g_X}$ 
    obtained by the ATLAS 2019 results~\cite{ATLAS:2019erb} (the CMS 2019 results~\cite{CMS:2019tbu}). 
We also show the upper bound (lighter red horizontal solid line) 
    for the narrow width approximation to be valid, where we have set $\Gamma_{Z^\prime} / m_{Z^\prime} < 3\%$.
The black solid (dashed) line shows the 95\% C.L. (68\% C.L.) BBN bound on $\Delta N_\mathrm{eff}$. 
The green-shaded region satisfies all constraints.
These three constraints are complementary to narrow down the allowed region
   to be $m_{Z^\prime} \gtrsim 4.5$ TeV and 
   $0.005 \lesssim \alpha_{g_X} \lesssim 0.05$.

In Fig.~\ref{Fig:xH_alpha}, we show allowed parameter regions in the ($x_H$, $\alpha_{g_X}$)-plain 
   for various $m_{Z^\prime}$ values. 
Fig.~\ref{Fig:xH_alpha}\subref{Fig:xH_alpha_mzp=5250} is for $m_{Z^\prime}=5.25$ TeV. 
The (blue) convex-downward solid line shows the lower bound on $\alpha_{g_X}$ as a function of $x_H$ 
    to reproduce the observed DM relic density. 
The (red) convex-upward solid (dashed) line shows the upper bound on $\alpha_{g_X}$ 
    obtained from the search results for $Z^\prime$ boson resonance 
    by the ATLAS~\cite{ATLAS:2019erb} (CMS~\cite{CMS:2019tbu}) Collaboration,
    along with the upper bound (lighter red horizontal solid line) 
    for which the narrow width approximation is valid (we have set $\Gamma_{Z^\prime} / m_{Z^\prime} \leq 3\%$).
The green shaded region satisfies all constraints.
These three constraints are complementary to narrow down the allowed region
   to be $-0.8 \lesssim x_H \lesssim -0.2$ and 
   $0.01 \lesssim \alpha_{g_X} \lesssim 0.03$.
Figs.~\ref{Fig:xH_alpha}\subref{Fig:xH_alpha_mzp=5000},
   \ref{Fig:xH_alpha}\subref{Fig:xH_alpha_mzp=4750}
   and \ref{Fig:xH_alpha}\subref{Fig:xH_alpha_mzp=4500}
   are the same as Fig.~\ref{Fig:xH_alpha}\subref{Fig:xH_alpha_mzp=5250},
   but $m_{Z^\prime}=5$ TeV, 4.75 TeV and 4.5 TeV, respectively.
From Fig.~\ref{Fig:xH_alpha}\subref{Fig:xH_alpha_mzp=4500},
   the allowed region to satisfy these three constraints indicates
   $-0.65 \lesssim x_H \lesssim -0.45$ and $0.008 \lesssim \alpha_{g_X} \lesssim 0.009$
   for fixed $m_{Z^\prime}=4.5$ TeV.
As $m_{Z^\prime}$ decreases, the LHC upper bound lines are shifted downward,
   while the DM lower bound line remains almost the same (it slightly moves to downward).
Therefore, the allowed region between the LHC upper bounds and the DM lower bound narrows.
On the other hand, the BBN bound remains almost the same, 
  so that the green shaded region disappears for $m_{Z^\prime} < 4.5$ TeV,
  as shown in Fig.~\ref{Fig:xH_alpha}\subref{Fig:xH_alpha_mzp=4500}.

%%%%%%%%%%%%%%%%%%%%%%%%%%%%%%
%\vspace{10mm}
\section{Determining the number of RHNs at High-Luminosity LHC}
%%%%%%%%%%%%%%%%%%%%%%%%%%%%%%

The SM neutrinos are Dirac particles in our model.  
This is in sharp contrast with usual U(1)$_X$ extension of the SM, where RHNs are heavy Majorana states. 
Since the RHNs are singlet under the SM gauge groups and 
   the Dirac Yukawa coupling constants are very small to reproduce the neutirno oscillation data,  
   the $Z^\prime$ boson interaction is the only way for RHNs to communicate with the SM particles. 
The ATLAS and CMS Collaborations continue to search for $Z^\prime$ boson resonance. 
Once discovered, the $Z^\prime$ boson will allows us to investigate physics of the RHNs 
   through precise measurements of $Z^\prime$ boson properties. 
In this section, we discuss an implication of the Dirac neutrinos in our model to LHC physics.

If the RHNs are heavy Majorana particles as in the minimal U(1)$_X$ model, 
  a pair of RHNs, if kinematically allowed, can be produced by $Z^\prime$ boson decay
  once $Z^\prime$ boson is produced at the LHC.  
A produced RHN subsequently decays to weak gauge bosons/Higgs boson plus lepton. 
Because of the Majorana nature of the RHN, the final states include same-sign leptons. 
This is the ``smoking-gun" signature of the lepton number violation, and we can expect
  a high possibility to detect such final states with much less SM background.  
For studies of such a signature at the LHC, see, for example, 
 Refs.~\cite{Huitu:2008gf, Kang:2015uoc, Cox:2017eme, Accomando:2017qcs, Das:2017flq, Das:2017deo}. 
See also Refs.~\cite{Das:2018tbd, Nakajima:2022pkd} for studies of RHN production at the International Linear Collider.

The Majorana RHNs are heavy and can be produced only if they are kinematically allowed, 
    while the Dirac neutrinos in our model are always included in the $Z^\prime$ boson decay products. 
Although, like the SM left-handed neutrinos, RHNs produced at colliders are hard to detect directly, 
  there is a possibility to measure them indirectly. 
Recall the great success of the LEP experiments that precisely measured the $Z$ boson decay width 
    and determined the number of the SM (left-handed) neutrinos to be three \cite{ALEPH:2005ab}. 
We notice that the $Z^\prime$ production is analogous to the $Z$ production at the LEP. 
Although the RHNs produced by the $Z^\prime$ boson decay will be undetectable, 
    the total $Z^\prime$ boson decay width carries the information of the invisible decay width.  
A precise measurement of the $Z^\prime$ boson cross section at the LHC may reveal the existence of RHNs. 
To illustrate this idea, we calculate the differential the cross section for the process 
    with the dilepton final state,  
    $pp \to \ell^+ \ell^- +X$ ($\ell=e$ or $\mu$) mediated by photon, $Z$ boson and $Z^\prime$ boson 
    at the LHC with a collider energy $\sqrt{s}=14$ TeV.

%%%%%%%%%%%%%%%%%%%%%%%%%%%%%%%%%%%%%%%%%%%%%%%%%
\begin{figure}[t]
\begin{center}
\includegraphics[scale=0.9]{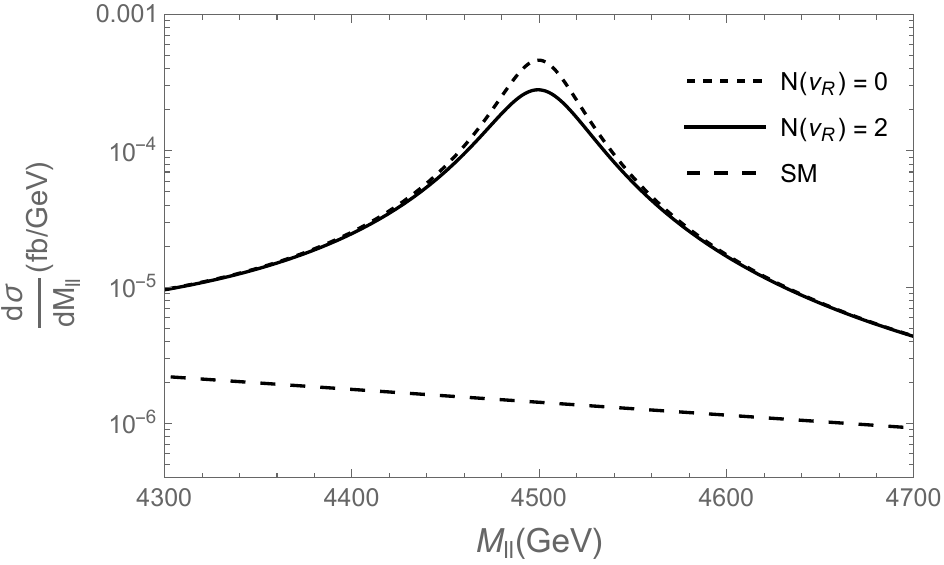}
%{\includegraphics*[width=1\linewidth]{FigLHC.eps}}
\caption{
The differential cross section for $pp \to \ell^+ \ell^- +  X $ 
  at the High-Luminosity LHC with $\sqrt{s}=14$ TeV for $m_{Z^\prime}=4.5$ TeV, $\alpha_{g_X}=0.008$, and $x_H=-0.5$. 
The solid and dashed curves correspond to the results for $N(\nu_R)=2$ and $0$, respectively.  
The horizontal long-dashed line represents the SM cross section, 
  which is negligibly small compared with the $Z^\prime$ boson mediated process. 
}
\label{FigLHC}
\end{center}
\end{figure}
%%%%%%%%%%%%%%%%%%%%%%%%%%%%%%%%%%%%%%%%%%%%%%%%%

For the $Z^\prime$ boson mediated process, we consider two cases, 
   $N(\nu_R)=0$ and $N(\nu_R)=2$, where $N(\nu_R)$ is the number of light RHNs.
The case $N(\nu_R)=0$ corresponds to the minimal U(1)$_X$ model with all RHN Majorana neutrino masses $> m_{Z^\prime}/2$. 
The total $Z^\prime$ boson decay width given in Eq.~(\ref{Eq:DecayWidthZp}) corresponds to $N(\nu_R)=2$ 
   while we add $-2$ to the inside of the bracket on the right-hand side for $N(\nu_R)=0$. 
Fig.~\ref{FigLHC} shows the differential cross sections for $pp \to \ell^+\ell^- + X$ ($\ell=e$ or $\mu$) 
   for $m_{Z^\prime}=4.5$ TeV, $\alpha_{g_X}=0.008$, and $x_H=-0.5$,
   along with the SM cross section mediated by the $Z$-boson and photon (horizontal long-dashed line). 
The solid and dashed curves correspond to the results for $N(\nu_R)=2$ and $0$, respectively.  
The dependence of the total decay width on $N(\nu_R)$ reflects the resultant cross sections
   (width and height at the peak of the cross section). 
When we choose a kinematical region for the invariant mass of the dilepton final state 
  to be in the range of $m_{Z^\prime}- 100  \leq m_{ll}[{\rm GeV}] \leq  M_{Z^\prime} + 100$, for example,   
   the signal events of $ 60.1$ and $80.3$ for $N(\nu_R)=2$ and $0$, respectively, 
   would be observed with the goal integrated luminosity of 3000/fb at the High-Luminosity LHC.   
The difference between $N(\nu_R)=2$ and $0$ are distinguishable with about $2.6 \sigma$ significance.   
If we simply combine the events for the final di-electon and di-muon, 
   the signal events become $120.2$ and $160.6$ for $N(\nu_R)=2$ and $0$, respectively, 
   the significant increases to $3.7 \sigma$.  

%%%%%%%%%%%%%%%%%%%%%%%%%%%%%%
%\vspace{10mm}
\section{Conclusions and Discussions}
%%%%%%%%%%%%%%%%%%%%%%%%%%%%%%

We have proposed a minimal SUSY U(1)$_X$ model with conserved R-parity. 
The gauge group U(1)$_X$ is a generalization of the well-known U(1)$_{B-L}$,
  and the U(1)$_X$ charge of an MSSM field is assigned as a linear combination of its $B-L$ charge
  and U(1) hypercharge, $Q^f_X = x_H \, Q_Y^f + Q_{B-L}^f$, with a common parameter $x_H$. 
Three right-handed neutrino chiral superfields are introduced, which make the model free 
  from all gauge and mixed gauge-gravitational anomalies, 
  but no U(1)$_X$ Higgs fields with U(1)$_X$ charge $\pm 2$ are included in the particle content. 
We assign an even R-parity to one right-handed neutrino superfield $\Psi$, 
   while an odd parity for the other two right-handed neutrino superfields, as usual. 
The scalar component of $\Psi$ ($\phi$) plays a role of the U(1)$_X$ Higgs field to break the U(1)$_X$ gauge symmetry.   
We have shown that the negative mass squared for $\phi$ is radiatively generated by the RG evolution, 
and hence the U(1)$_X$ symmetry is radiatively broken. 
Due to the absence of U(1)$_X$ Higgs fields with U(1)$_X$ charge $\pm 2$ and our novel R-parity assignment, 
   three light neutrinos consist of one massless neutrino and two massive Dirac neutrinos. 
Since R-parity is conserved, the LSP neutralino is a prime candidate for the dark matter. 
Our model offers a new dark matter candidate ($\chi_\ell$) which is 
  a linear combination of the fermion component of $\Psi$ and the U(1)$_X$ gaugino. 
Assuming this $\chi_\ell$ is the LSP, we have investigated the dark matter phenomenology 
 to find the parameter region that reproduces the observed dark matter relic density. 
We have found that an enhancement of dark matter pair annihilation process through $Z^\prime$ boson 
 resonance in the $s$-channel is crucial for reproducing the observed dark matter density 
 and hence the dark matter mass is close to half of the $Z^\prime$ boson mass.  
As a result, the dark matter relic density constrains three parameters of the model, 
 $\alpha_{g_X}$, $m_{Z^\prime}$ and $x_H$.  
The $Z^\prime$ boson resonance has been searched for at the LHC Ran-2. 
Using the final LHC Run-2 results reported by the ATLAS and CMS Collaborations,  
  we have derived the constraints on the three parameters, $\alpha_{g_X}$, $m_{Z^\prime}$ and $x_H$. 
In the early universe, two right-handed components of the light Dirac neutrinos are in thermal equilibrium 
  with the Standard Model particle plasma through the U(1)$_X$ gauge interaction. 
Their energy density increases the expansion rate of the universe at the Big Bang Nucleosynthesis era. 
To avoid disrupting the success of the Big Bang Nucleosynthesis, 
  we have found the lower bound on the scale of U(1)$_X$ symmetry breaking, $v_X = m_{Z^\prime}/g_X$, 
  for a fixed value of $x_H$. 
Combining three constraints obtained from the observed dark matter relic density, 
  the $Z^\prime$ boson resonance search at the LHC Run-2 and the success of Big Bang Nucleosynthesis, 
  we have narrowed down the allowed parameter region. 
We have also discussed a possibility of determining the number of light right-handed neutrinos 
  at the High-Luminosity LHC. 
The $Z^\prime$ boson, once discovered at the LHC, can be a probe for this purpose,  
 since the shape of the $Z^\prime$ boson resonance as a function of the invariant mass of the final state dilepton 
 carries the information of the number of light right-handed neutrinos.

Since the light neutrinos are Dirac particles in our model, their Dirac Yukawa couplings must be extremely small
  (${\cal O}(10^{-12})$) to fit the neutrino oscillation data. 
We need to consider a natural way of realizing such extremely small Yukawa couplings.  
To solve this problem, we may extend our model in 4-dimension to the brane-world framework 
   with 5-dimensional warped space-time~\cite{Randall:1999ee}.   
Arranging the bulk mass parameters for the bulk hypermultiplets corresponding to matter and Higgs fields in our model, 
   we can obtain a large hierarchy among parameters in 4-dimensional effective theory with mildly hierarchical model parameters in the original 5-dimensional theory. 
For example, we may arrange the bulk mass parameters for R-parity even right-handed neutrino superfields 
  to localize around one brane, while for the lepton doublet superfileds to localize around the other brane.
In this way, the 4-dimensional effective Yukawa coupling can be exponentially suppressed. 
For such an idea, see, for example, Ref.~\cite{Grossman:1999ra}.

%%%%%%%%%%%%%%%%%%%%%%%%%%%%%%
\vspace{10mm}
\section*{Acknowledgements}
%%%%%%%%%%%%%%%%%%%%%%%%%%%%%%

The work of N.O. is supported in part by the United States Department of Energy grant (DE-SC0012447).
S.O.~and D.-s.T.~are supported by the mathematical and
theoretical physics unit (Hikami unit) of the Okinawa Institute of
Science and Technology Graduate University. S.O.~and D.-s.T.~are also
supported by Japan Society for the Promotion of Science (JSPS),
Grant-in-Aid for Scientific Research (C), Grant Number JP18K03661.

%%%%%%%%%%%%%%%%%%%%%%%%%%%%%%%%%%

%%%%%%%%%%%%%%%%%%%%%%%%%%%%%%%%%%

%%%%%%%%%%%%%%%%%%%%%%%%%%%%%%
%\vspace{10mm}
%\bibliographystyle{plain}
%\bibliography{list}
%%%%%%%%%%%%%%%%%%%%%%%%%%%%%%

%%%%%%%%%%
%%%%%%%%%%
%%%%%%%%%%
%%%%%%%%%%
\end{document}